# Nonreciprocal Transmission in Nonlinear PT-Symmetric Metamaterials Using Epsilon-near-Zero Media Doped with Defects

Boyuan Jin and Christos Argyropoulos[*]
Department of Electrical and Computer Engineering, University of Nebraska-Lincoln,
Lincoln, Nebraska 68588, USA
*christos.argyropoulos@unl.edu

**Abstract**

Nonreciprocal transmission forms the basic operation mechanism of optical diodes and isolators and requires the tantalizing task of breaking the Lorentz reciprocity law. In this work, strong nonreciprocal transmission is demonstrated by using a compact nonlinear parity-time (PT) symmetric system based on epsilon-near-zero (ENZ) materials photonically doped with gain and loss defects and separated by an ultrathin air gap. The nonlinear response of this scalable configuration is triggered at relatively low optical intensities due to the strong electric field confinement in the defects. The extreme asymmetric field distribution achieved upon excitation from opposite incident directions, combined with the enhanced nonlinear properties of the proposed system, result in a pronounced self-induced nonreciprocal transmission. Cascade configurations with optimized geometrical dimensions are used to achieve self-induced nonreciprocal transmission with a maximum contrast, ideal for the design of new all-optical diodes. The presented robust nonreciprocal response occurs by operating at a frequency slightly shifted off the exceptional point but without breaking the PT-symmetric phase, different compared to previous works. The findings of this work can have a plethora of applications, such as nonreciprocal ultrathin coatings for the protection of sources or other sensitive equipment from external pulsed signals, circulators, and isolators.

## 1. Introduction

The concept of parity-time (PT)-symmetry in optics and photonics was inspired from the intriguing physics of quantum mechanics.[1] It can be achieved by using asymmetric configurations with balanced gain and loss parameters or, equivalently, refractive indices satisfying the relation $n(\vec{r}) = n^*(-\vec{r})$.[2-4] PT-symmetric photonic structures have revealed exotic and interesting optical functionalities, such as coherent perfect absorption and lasing, asymmetric diffraction, unidirectional invisibility, nonreciprocal light propagation, and extraordinary nonlinear effects.[2-23] In these systems, a special frequency point exists, named exceptional point (EP), where the two eigenvalues of the system merge and the effective non-Hermitian Hamiltonian becomes defective.[24, 25] The EP is the critical state found exactly before the system experiences a PT-breaking transition[24, 26-31] leading to larger than one transmission and eventually to spectral singularities or lasing. It has been demonstrated that the reflection becomes asymmetric at the EP by illuminating from opposite directions, while the transmittance is equal to unity due to reciprocity.[14] PT-symmetric systems have recently been designed by using different photonic and optical metamaterial schemes, such as zero index metamaterials (ZIMs), epsilon-near-zero (ENZ) materials, hyperbolic metamaterials, and multilayer Bragg gratings.[14, 17, 18, 32, 33]

Optical nonlinearity is an inevitable phenomenon in the material response when the input power of the incident radiation is increased.[34-37] By accounting the effect of nonlinearity into PT-symmetric devices operating at the broken PT-symmetric phase, away from the EP, where a spectral singularity exists, or other passive asymmetric systems, the transmission can become nonreciprocal and the Lorentz reciprocity law is broken.[38-45] The asymmetric transmission is of great importance to many photonic applications, such as optical diodes, isolators, and cloaks.[11, 19, 24] However, passive nonlinear nonreciprocal systems suffer from fundamental physical bounds between their nonreciprocal transmission contrast and asymmetric field distribution.[41, 43, 46, 47] In addition, lasing or other detrimental instabilities can occur for nonlinear PT-symmetric devices operating at the broken PT-symmetric phase that will limit their nonreciprocal response even in the presence of saturable nonlinearities.[40]

In this work, the aforementioned limitations on nonlinear self-induced nonreciprocity are broken by using a scalable and compact PT-symmetric active metamaterial constructed by two ENZ media separated by an ultrathin gap filled with air and photonically doped with gain and loss defects.[48] We demonstrate that the extremely asymmetric field distribution, combined with the enhanced nonlinear effects inside both defects, will lead to almost perfect nonreciprocal transmission with a pronounced contrast. The permeability of the two ENZ media also has a real part equal to zero in this configuration, due to the defects, leading to a zero refractive index operation that is matched to the surrounding free space.[49] Based on the proposed active metamaterial, we demonstrate strong nonreciprocal transmission close to the EP, as the input intensity of the incident waves launched from opposite directions is increased, but without breaking the PT-symmetric phase condition leading to larger than one transmission values or spectral singularities. The electric field is tightly confined and enhanced inside the photonic doping defect regions, triggering third-order nonlinear effects with relative low input intensity values.[50-53] The periodicity of the structure is further optimized to achieve complete transmission in the backward direction, and approximately zero transmission in the forward direction, in the case of cascading the proposed nonlinear PT-symmetric metamaterials. The presented strong self-induced nonreciprocity effect does not require magnetic-bias or other time-modulated schemes that can be difficult to be implemented, especially in the case of planar compact configurations and for higher frequencies.[44] Hence, it can be ideal for the design of new on-chip photonic devices, such as all-optical diodes, isolators, circulators, and protective layers for sensitive optical components.

## 2. Linear operation and EP formation

The geometry of the proposed PT-symmetric system is shown in Figure 1. Two identical ENZ media are separated by an ultrathin air gap, and a cylindrical defect with gain or loss is located at the center of each ENZ segment. In the following, the dimensions of the proposed structure are normalized to the wavelength $\lambda_{ENZ}$ where the permittivity of the ENZ medium becomes equal to zero. Hence, the presented results are general and can be obtained in different frequency ranges depending on the material availability. Materials or artificially engineered structures that exhibit ENZ response with minimum loss at different frequency ranges, such as microwaves,[48] mid-infrared (IR),[54, 55] and near-IR/visible,[37, 56-58] have been widely investigated and demonstrated. Although materials with loss are ubiquitous, active materials that exhibit gain are more difficult to find. However, several materials or configurations with strong gain coefficients have been recently proposed at microwave,[59, 60] mid-IR,[61] and near-IR/visible[62] frequencies. In the currently proposed structure, the real parts of each defect's permittivity are equal, while the imaginary parts have opposite signs but equal and small absolute values in order to obey the PT-symmetry condition but avoid detrimental stimulated emission effects. We assume the relative permittivity of the ENZ medium to be very close to zero $\varepsilon_{ENZ} = 0.0001$. The complex relative permittivity of the lossy defect is $\varepsilon_{Loss} = \varepsilon_d = 4 - i\delta$, where $\delta$ is an arbitrary real coefficient with a very small value, as it will be shown later. The active defect has a complex conjugate relative permittivity equal to $\varepsilon_{Gain} = \varepsilon_{Loss}^* = \varepsilon_d^* = 4 + i\delta$ in order to obey the PT-symmetric condition. The ENZ media and defects are non-magnetic and their relative permeabilities are $\mu_{ENZ} = \mu_d = 1$.

The propagation directions of the forward and backward input waves are defined in Figure 1. The currently used excitation is a normal incident plane wave polarized along the y-axis with arbitrary wavelength $\lambda$ and variable input intensity $I_0$. We set the horizontal boundaries in our simulations[63] (along and parallel to the x-z plane) to be perfect electric conductors (PEC), while the input electric field is polarized perpendicular to the x-z plane. Therefore, the structure shown in Figure 1 can be considered either as a single waveguide with width $a$ terminated by metallic lateral boundary surfaces or, equivalently, the unit cell of a periodic structure along the y-axis made of an array of waveguides with period $a$. The out-of-plane dimension of the structure along the z-axis is assumed to be much larger than the wavelength $\lambda$ of the incident radiation. Hence, the proposed structure can be treated as two-dimensional (2D) or, alternatively, infinite and uniform in the z-direction.

We start our investigation from the linear characterization of this structure, similar to the simple scenario of very low input intensity $I_0$ illumination, when nonlinear effects are not expected to be excited and can be neglected. The dimensions are selected to be $a = \lambda_{ENZ}$, $b = 3.265\lambda_{ENZ}$, $r = 0.194\lambda_{ENZ}$, and $d = 0.25\lambda_{ENZ}$, all of them normalized to $\lambda_{ENZ}$, the wavelength where the medium hosting the defects has almost zero permittivity. It will be demonstrated that the transmittance can reach one in this configuration, while the reflectance is equal to zero only for one direction (unidirectional reflectionless transparency), at a particular frequency point, known as EP.[64] This property, combined with the resulted asymmetric field distribution and nonlinear effects for higher input intensities, will lead to strong self-induced nonreciprocal response. The radius $(r)$ value of each defect is chosen in order to obtain zero refractive index collective response and matching to the surrounding free space. Finally, we assume that the dispersion of the defect materials is negligible when the operation wavelength is close to $\lambda_{ENZ}$.

The scattering matrix S of the proposed structure is extracted from full-wave linear simulations.[63] The transfer matrix M can be derived by the computed S matrix: [31, 65]

$$M = \begin{bmatrix} M_{11} & M_{12} \\ M_{21} & M_{22} \end{bmatrix} = \begin{bmatrix} S_{21} - \dfrac{S_{22}S_{11}}{S_{12}} & \dfrac{S_{22}}{S_{12}} \\ -\dfrac{S_{11}}{S_{12}} & \dfrac{1}{S_{12}} \end{bmatrix}. \quad (1)$$

The eigenvalues of the matrix M are calculated by: [31, 66]

$$\eta_{1,2} = \frac{M_{11} + M_{22}}{2} \pm \sqrt{\left(\frac{M_{11} + M_{22}}{2}\right)^2 - 1} \quad (2)$$

We plot in Figure 2(a) the evolution of the two complex eigenvalues $\eta_{1,2}$ at the $\lambda_{ENZ}$ wavelength as a function of a small variation in the gain/loss parameter $\delta$ around $\delta_0 = 0.01$. Interestingly, the two eigenvalues coincide on the unit circle only for the very small value of $\delta = \delta_0 = 0.01$, which is a clear indication of an EP obtained at the ENZ frequency of the proposed system.[18] The EP formation is also verified by computing the reflectance and transmittance spectra around the ENZ wavelength with results shown in Figures 2(b) and 2(c), respectively. The forward and backward transmittances are equal to one $(T_F = T_B = 1)$ at $\lambda = \lambda_{ENZ}$ (EP), obeying the reciprocity condition, while the forward reflectance is $R_F = 15.8$ (not surprisingly larger than one due to the gain material) and the backward reflectance is $R_B = 0$ leading to unidirectional reflectionless transparency, a typical property of EPs.[25]

Note that we choose an ultrathin air gap to separate the two ENZ media ($d = 0.25\lambda_{ENZ}$). This is the minimum distance in order to obtain an EP in the currently studied active photonic metamaterial, as it is explained in more detail in the Supplemental Material.[67] Interestingly, our simulations demonstrate that the EP can happen for particular distances $d$ of the air spacer layer, which is different from previous relevant works of other PT-symmetric systems.[68, 69] where the EP was found to be independent of the distance between the loss and gain parts. The impedances of the zero refractive index loss/active media in the current configuration have a reactive or imaginary part due to the stored reactive energy along their elongated thicknesses, different from the purely real impedances used in previous works.[68, 69] Hence, the EP can be tuned as we vary the length of the spacer layer between the two ENZ media, as it is demonstrated in the Supplemental Material.[67] It is also worth noting that the thickness of each ENZ media $(b = 3.265\lambda_{ENZ})$ has this particular value in order to achieve the EP at the $\lambda = \lambda_{ENZ}$ wavelength for a very small value of the gain/loss coefficient, i.e., $\delta_0 = 0.01$. However, the length $b$ can be further reduced, making the structure even more compact, if we increase the value of $\delta_0$ without affecting the EP formation at $\lambda = \lambda_{ENZ}$.[18] The electric field distribution at the EP is also computed and shown in Figure 2(d). It is found to be strongly asymmetric and enhanced inside the defects. This is the main physical mechanism that will generate the self-induced nonreciprocity when nonlinearities will be triggered.

### 3. Self-induced nonreciprocal transmission

When the input intensity of the incident radiation is increased, third-order nonlinear effects will be triggered, especially along the defects where the field enhancement is maximum, as it was shown in Figure 2(d). We assume that the defects have a relatively low third-order nonlinear susceptibility equal to $\chi^{(3)} = 6 \times 10^{-20}$ m²/V², consisting a typical value of several dielectric materials.[34, 44] The nonlinearity of the ENZ media is currently neglected, because the field enhancement intensity in the ENZ media is much smaller compared to the field in the defects, clearly depicted in Figure 2(d). However, we have verified in the Supplemental Material that the significant nonreciprocal transmission can be further improved and achieved with even lower input intensity values, when we take into account in our calculations the nonlinear susceptibility $\chi^{(3)}_{ENZ}$ of the ENZ media, which is also found to generate a perceivable Kerr effect.[67] The third-order nonlinear Kerr effect is expected to dominate in the proposed configuration under high intensity illumination. It will introduce a nonlinear polarization given by $P_{NL} = \varepsilon_0 \chi^{(3)} |E|^2 E$ in the presented nonlinear active metamaterial, where $\varepsilon_0$ is the permittivity of free space and $E$ is the local electric field distribution.[34] Consequently, this polarization will lead to a modulation in the permittivity given by: $\varepsilon = \varepsilon_L + \chi^{(3)} |E|^2$, where $\varepsilon_L$ is the complex linear permittivity of the loss and gain defects given in the previous section.[34, 70-72]

As the input intensity $I_0$ is increased, the transmittance from both sides will decrease monotonously due to detuning based on the third-order Kerr nonlinear effect that will mainly affect the matching condition of this structure. This effect is clearly depicted in Figure 3, where we plot the variation of transmittance from both illumination directions at $\lambda = \lambda_{ENZ}$ and as a function of the input intensity. Obviously, the system's response becomes tunable as the input intensity is increased. The active metamaterial does not operate anymore exactly at the EP as we increase the input intensity, interestingly, without breaking the PT-symmetric phase condition (i.e., transmission is always kept below one). This effect is different compared to other PT-symmetric systems based on Bragg scattering, where it was found that the EP remains unaffected by the presence of Kerr nonlinearities.[73] When the input intensity has very low values, less than 10 kW/cm$^2$, the influence of nonlinear effects is extremely weak, and the system operates approximately in the linear regime, i.e., the structure is totally transparent from both illumination directions and the transmission at the EP is reciprocal. However, the transmittance values rapidly decrease when the illumination intensity $I_0$ is increased from 100 kW/cm$^2$ to 1 GW/cm$^2$. Interestingly, the forward transmittance declines faster than the backward, a direct indication of self-induced nonreciprocity triggered by nonlinear effects. The stronger nonlinear response felt by the forward incident wave is due to the robust electric field enhancement obtained along the defect loaded with the active material, as was shown before in Figure 2(d), which is located first along the forward propagation path. The difference between the forward and backward transmittance is also shown in Figure 3. This metric clearly demonstrates the obtained nonreciprocal transmission and its contrast. More precisely, the difference in transmittance ($T_B - T_F$) at very low input intensity values ($I_0 = 5.62$ MW/cm$^2$) reaches its maximum value 0.53, where $T_F = 0.35$ and $T_B = 0.88$. This is a decent nonreciprocal response, which is comparable to other relevant works.[22, 41, 74] However, our goal is to substantially improve the self-induced nonreciprocity in order to make the proposed structure more appropriate for the design of optical diodes, isolators, circulators, and protective layers for sensitive optical components. Note that after reaching the threshold high input intensity value of 1 GW/cm$^2$, the transmittance from both sides becomes small and saturates to zero, similar to the saturable nonlinearity effect.[40]

In order to further enhance the nonreciprocal response of the system and achieve increased difference between forward and backward transmittances, the periodicity of the proposed PT-symmetric nonlinear system is optimized. The resulted optimized structure will operate slightly off the EP (or ENZ wavelength), however without breaking the PT-symmetric phase. The contour plots shown in Figure 4 demonstrate the forward and backward transmittances and their difference for a range of input intensity and periodicity $a$ values. The input intensity range is chosen to vary between the moderate values of 0.1 MW/cm$^2$ and 100 MW/cm$^2$, where the largest transmittance change is achieved. Both forward and backward transmittances greatly fluctuate in this range and exhibit an abrupt jump from close to zero to approximately one values for lower (forward transmittance) or higher (backward transmittance) input intensities, as it can be seen in Figures 4(a) and 4(b). The transmittance can slightly exceed one at some points due to the active gain material used in this system. Interestingly, the peak of forward transmittance occurs at low input intensities, while the peak of backward transmittance is obtained at relatively higher intensity values. Our ultimate goal to achieve perfect self-induced nonreciprocal response requires the transmittance at one direction to be very low (ideally zero), while the transmittance at the other direction to approach one (perfect transparency). Therefore, we need to select a periodicity value that is as close as possible to the maximum transmission point in one graph while approaches the minimum point in the other graph. Note that the periodicity values affect both the impedance matching and EP formation, or equivalently the real and imaginary parts of the effective permeability of the proposed ENZ doped with defect metamaterial.[14, 18, 49] The point marked by the star symbol in Figure 4 is an ideal option to achieve maximum nonreciprocal transmission contrast, as demonstrated in Figure 4(c), corresponding to $a = 0.891\lambda_{ENZ}$ and a low input intensity value of $I_0 = 73$ MW/cm². Stronger nonreciprocal transmission can be achieved at this point, which is slightly off the EP, compared to the results presented before in Figure 3 obtained exactly at the EP and also depicted by the triangle in Figure 4. Hence, by slightly deviating our operation wavelength compared to the EP of the PT-symmetric structure, which is equivalent to changing the periodicity of the structure, unity transmittance from one side and almost zero transmittance from the other side can be achieved due to the strong nonreciprocal interaction caused by the enhanced nonlinearity at the gain/loss defect asymmetric field regions.

The dependence of transmittance on the input intensity, when the period of the proposed optimized structure is $a = 0.891\lambda_{ENZ}$, is shown in Figure 5(a). Similar to Figure 3(a), when the input intensity is less than 1 MW/cm², the system is linear, and the transmission is reciprocal and very close to zero for both illumination sides. However, the shapes of these transmittance curves are significantly different compared to Figure 3(a). In this case, the transmittance from each direction first increases and then decreases, as we increase the input intensity of the incident wave. The forward transmittance peak is generally lower, and its sharp increase occurs at lower input intensities compared to the backward illumination case. The forward transmittance is rapidly decreased at higher input intensities. On the contrary, the backward transmission can reach full transparency (transmittance equal to one) for moderate input intensity values and with a large contrast compared to the forward illumination case. The lagging between the two sharp transmission peaks generates a large transmittance contrast when the structure is illuminated from opposite directions. The maximum difference in transmittance $T_B - T_F$ can reach 0.89 at low input intensity values $I_0 = 73$ MW/cm², as can be seen in Figure 5(b), where $T_F = 0.11$ and $T_B = 1$. Another striking feature shown in Figure 5(b) is that the transmittance difference can be either positive or negative depending on the input intensity values. At relatively low input intensities, the forward transmittance is larger, and the opposite is valid for high input intensities, making the self-induced nonreciprocal response tunable depending on the incident intensity values. The reflectance of this configuration is also shown in Figure 5(c). The reflectance is equal to zero at the unitary transmission peak in the backward direction, indicating that no power is reflected and unidirectional reflectionless transparency is achieved only from one direction, a distinct characteristic of EPs.[31] However, this nonlinear-induced EP response is obtained only at the backward direction in the current nonlinear PT-symmetric structure while the forward direction transmission for the same input intensity is very low and close to zero. The electric field distribution for $I_0 = 73$ MW/cm² is also presented in Figure 5(d). The electric field is concentrated and enhanced along the defect regions, as it is clearly shown by the full-wave nonlinear simulation results. Therefore, the nonlinear properties of the defects will dominate the nonlinear response of this system. However, the incorporation of the nonlinear ENZ properties will further improve the nonreciprocal response.[67] The drastic field enhancement can make the nonlinear effects to be triggered with lower intensity values, as it happens in the current structure. Note that the proposed configuration does not act as a cavity and the current field enhancement values, which are relative low [Figure 5(d)], combined with the low gain coefficient are not expected to make the active defect to reach the lasing or strong-coupling regime.

Figure 5 demonstrated that, at some particular input intensity value, the optical power can be completely transmitted in the backward direction but is approximately equal to zero in the forward direction, similar to a nonreciprocal waveguide system being in the cutoff regime only for the forward incident waves. This feature directly implies that the obtained nonreciprocal transmission can be further boosted by cascading multiple nonlinear PT-symmetric metamaterials with the same dimensions and properties compared to the structure with response presented in Figure 5. Figures 6(a) and 7(a) show the schematics of the proposed cascade nonreciprocal metamaterial systems composed of either two or three stages, respectively. When light propagates in the backward direction, the attenuation through each PT-symmetric system is very small due to the perfect transparency condition obtained at a particular input intensity value. As a result, the input power is maintained along the propagation through the first cascade level and can be used to turn transparent the next cascade stage via the Kerr nonlinearity mechanism described before. Besides, the reflection becomes nearly zero, as it can be seen in Figure 5(c), when the backward transmittance approaches to one (unidirectional reflectionless transparency) for the single cascade stage system. Therefore, the interference between the transmitted wave from one stage and the reflected wave from the next stage is extremely week. Thus, the optical power can be kept almost constant after the wave propagates through each cascade stage in the backward direction. On the contrary, the transmittance of the first PT-symmetric cascade system is only 0.11 for light propagating in the forward direction at $I_0 = 73$ MW/cm$^2$, which means that only approximately ten percent of the total input power can penetrate through the first stage. The transmittance at the forward direction is always relatively low when sweeping the input intensity in Figure 5(a), and the maximum peak value is only 0.49. Therefore, the optical power can be greatly attenuated after each cascade stage along the forward direction. This consecutive attenuation effect leads to a further substantial enhancement in the self-induced nonreciprocal transmission level when cascade configurations are used.

Figure 6(b) demonstrates the transmittance at different directions by cascading two identical PT-symmetric systems with schematic shown in Figure 6(a). The dimension parameters are the same as those used in Figure 5 except for a slightly detuning in the periodicity due to the increased structure length, leading to an optimized value $a = 0.928\lambda_{ENZ}$. In this case, the values of forward and backward transmittances at $\lambda = \lambda_{ENZ}$ for the 2-stage cascade system are 0.024 and 1, respectively, for a fixed input intensity of $I_0 = 44.5$ MW/cm$^2$, which, interestingly, is lower compared to the single stage device due to the longer length of the current nonlinear structure. The 2-stage cascade system is totally transparent to the backward direction illumination, while the forward transmittance is decreased by almost five times compared to the single stage PT-symmetric system shown in Figure 5(a). Moreover, the forward and backward transmittances of the 3-stage cascade system, with schematic depicted in Figure 7(a), are shown in Figure 7(b), where the periodicity is slightly larger compared to the 2-stage case $(a = 0.938\lambda_{ENZ})$. In this case, the backward transmittance remains one (perfect transparency) at $\lambda = \lambda_{ENZ}$, interestingly, for even lower input intensity values $I_0 = 38$ MW/cm$^2$ compared to the 2-stage configuration, while the forward transmittance is further decreased to 0.01. This extremely low forward transmittance value is almost zero. Hence, the presented 3-stage cascade active metamaterial can be an ideal structure to achieve an optical diode with low input intensity values. The transmittance difference or contrast $T_B - T_F$ for the 2- and 3-stage cascade systems is also shown in Figures 6(c) and 7(c), respectively, where the substantial change between forward and backward transmittance is clearly depicted. A more detailed comparison between the response of the single and multiple cascade PT-symmetric configurations can be found in the Supplemental Material.[67] Note that the nonreciprocal transmission contrast can be further improved if configurations with even more cascade layers are considered.

Although the proposed device exhibits an impressively strong self-induced nonreciprocal performance, especially in the cascade configuration, the degree of nonreciprocity will be limited to a range of input signal intensities when illuminated by one side and will suffer from the 'dynamic reciprocity' problem, meaning that the device cannot operate correctly if it is simultaneously excited from both directions.[75, 76] This inherent problem can be alleviated by using pulsed illumination. Hence, the proposed structure can indeed be used as an optical diode or filter to protect sources or other sensitive equipment from strong external pulsed signals.[47] However, it is interesting to investigate what would happen in the nonreciprocal response in the practical situation when two incident waves are simultaneously launched from the opposite directions, a prohibited regime due to the 'dynamic reciprocity' problem. Figure 8(a) shows a single-stage PT-symmetric active metamaterial with the same dimensions and parameters as those used to obtain the results presented in Figure 5. The input intensities from the left and right ports (ports 1 and 2) are $I_1$ and $I_2$, respectively. In the case of single illumination, it was demonstrated in Figure 5 that the nonreciprocal transmission can reach its maximum contrast for $I_0 = 73$ MW/cm$^2$. The corresponding backward and forward transmittances were $T_B = 1$ and $T_F = 0.11$, respectively, in this case. Next, we simultaneously excite the nonlinear structure from both directions, as it is shown in Figure 8(a). However, we keep the intensity of one input port to be fixed at 73 MW/cm$^2$, while varying the input intensity of the other port, and observe its effect on the collective nonreciprocal transmission. This is shown by the black solid curve in Figure 8(b), where the output power ratio $P_{out2}/(P_1 + P_2)$ is plotted as a function of $I_2$ in the case of fixed $I_1 = 73$ MW/cm$^2$. Note that $P_{out2}$ is the measured output power at port 2, whereas $P_1$ and $P_2$ are the input powers along ports 1 and 2, respectively. The red dashed curve in Figure 8(b) shows the output power ratio calculated from the other side $P_{out1}/(P_1+P_2)$ as a function of $I_1$ when the input intensity in this case is fixed at port 2 $I_2 = 73$ MW/cm$^2$.

It can be seen in Figure 8(b) that the nonreciprocal effect (large contrast between output powers at ports 1 and 2) is almost unaffected until we reach a threshold input intensity value at each output port with values approximately equal to 20 MW/cm$^2$. After this threshold value, the nonreciprocity would deteriorate and the PT-symmetric structure can no longer work as an isolator, in agreement to the 'dynamic reciprocity' prediction.[75, 76] However, it is interesting to note that when the additional input intensity at the output port is lower and approximately equal to 17.2 MW/cm$^2$, the self-induced nonreciprocal performance can be further improved and achieve a more pronounced transmittance contrast ratio, even for the non-cascade configuration. Around this point, shown by a dashed line in Figure 8(b), the backward and forward transmittances are $T_B = 1$ and $T_F = 0.0016$, respectively, and the transmittance contrast $T_B/T_F$ has a 70-fold improvement compared to the response achieved in Figure 5. Hence, even in the case of multiple excitation waves, there is an input intensity range that the device can still achieve strong self-induced nonreciprocity, which can be further improved for some particular combination of low input intensity values radiated from ports 1 or 2. Finally, note that the results presented in Figure 8(b) become identical to Figure 5 when the input intensity at the output port has very small values, similar to the single wave illumination scenario.

## 4. Conclusions

In this work, we presented strong self-induced nonreciprocal transmission by using new compact and scalable nonlinear PT-symmetric zero refractive index metamaterials. The asymmetric and strong field enhancement in the gain/loss defects led to a substantially boosting in the Kerr nonlinear effect, which caused the presented enhanced nonreciprocal response to be triggered by relative low input intensity values. The dependence of the nonreciprocal transmittance on the geometrical parameters, including periodicity and gap length, was also investigated, and the dimensions were optimized to maximize the nonreciprocal transmission. The presented effect occurred by operating at a frequency slightly shifted off the EP but without breaking the PT-symmetric phase. We also demonstrated that the nonreciprocal transmission contrast can be further enhanced by cascading multiple PT-symmetric metamaterials or using simultaneous excitations from both sides. Finally, note that the proposed structure was found to be able to achieve EP formation and strong self-induced nonreciprocal transmission even in the case of increased losses in the ENZ media.[67]

Strong nonreciprocal isolation contrast combined with large and asymmetric fields are obtained with the proposed active nonlinear structure even in the case of unitary transmission from one side, a performance not restrained by the limitations in transmission of passive asymmetric structures.[77] Hence, the current work extends and breaks the limits of the passive nonlinear nonreciprocal systems by incorporating active (gain) media.[46] To sum up, the proposed scalable PT-symmetric system can indeed operate as an all-optical self-induced isolator or diode for pulsed illumination without requiring any outside bias to achieve isolation (i.e., magnetism, time-modulation). It can pave the way to the design of new compact on-chip nonreciprocal photonic devices, such as topological waveguides or circulators. We expect our theoretical findings to be experimentally tested using different realistic platforms due to the scalability of the proposed concept in different frequency regimes, where several materials exist with a broad range of gain, loss, and nonlinear parameters. The findings of this work connect the fields of PT-symmetric ENZ metamaterials and nonlinear optics to the emerging research area of bias-free self-induced nonreciprocal photonics.

**Acknowledgements**
This work was partially supported by the NSF (Grant No. DMR-1709612), the NSF Nebraska Materials Research Science and Engineering Center (Grant No. DMR-1420645), Office of Naval Research Young Investigator Program (ONR-YIP) Award (Grant No. N00014-19-1-2384), and NSF-Nebraska-EPSCoR (Grant No. OIA-1557417).

**Figures**

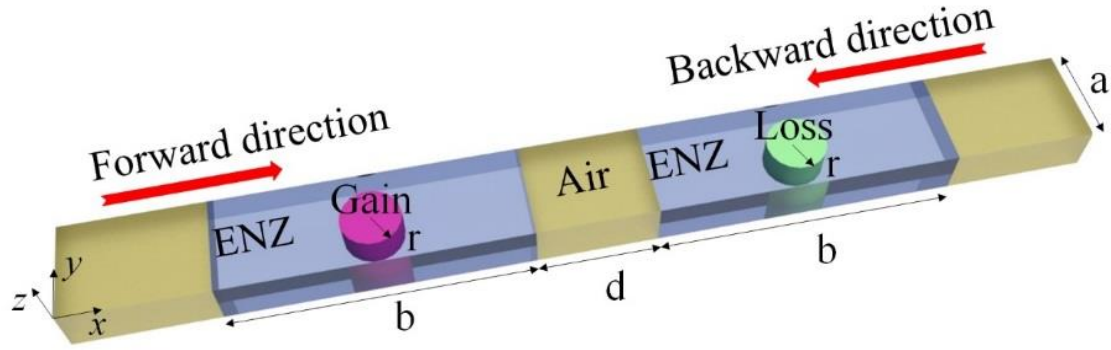

**Figure 1.** Schematic of the proposed PT-symmetric structure composed of two identical ENZ media loaded with gain and loss defects and separated by an ultrathin air gap.

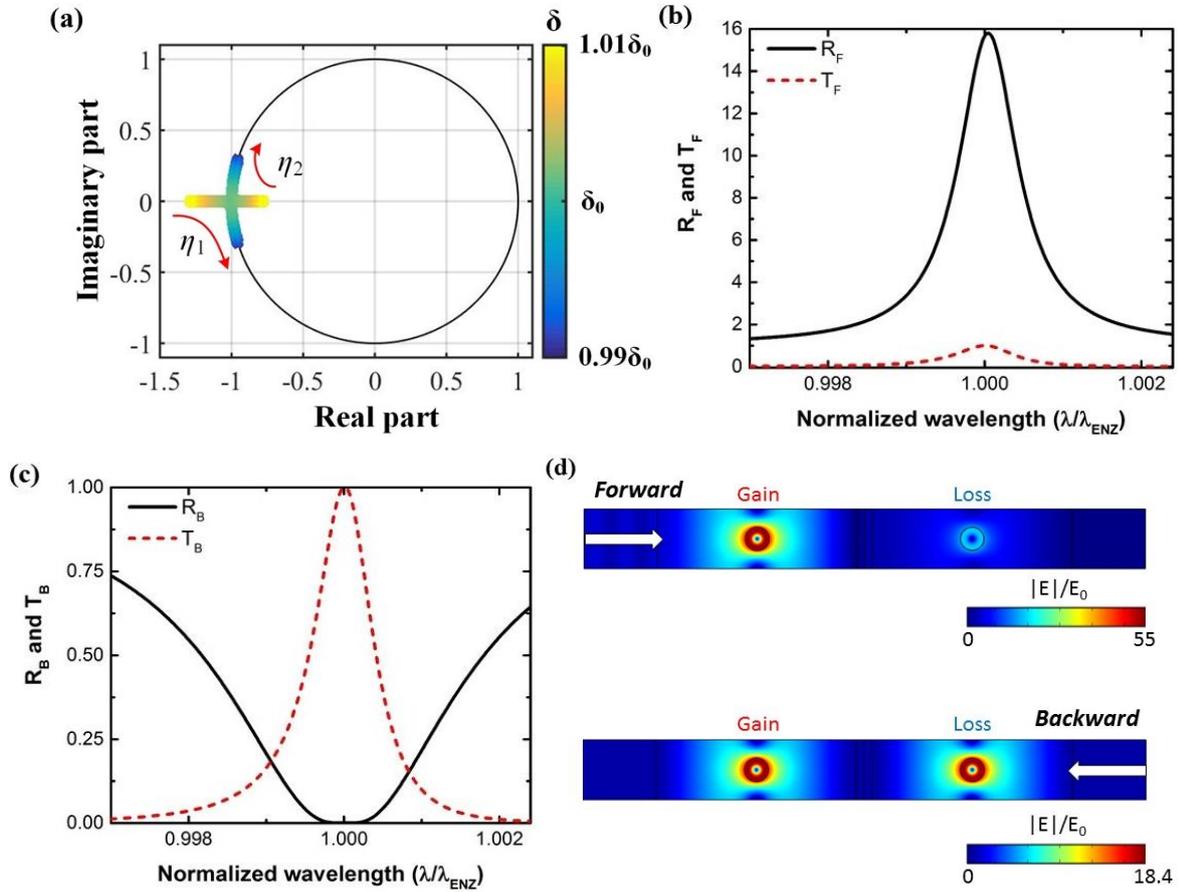

**Figure 2.** (a) Evolutions of the complex eigenvalues $\eta_1$ and $\eta_2$ as a function of the defect's permittivity imaginary part. The computed spectra of the linear transmittance $(T)$ and reflectance $(R)$ for (b) forward or (c) backward illumination clearly demonstrate the formation of an EP at the wavelength $\lambda = \lambda_{ENZ}$. (d) The electric field distribution at the EP under forward and backward illumination.

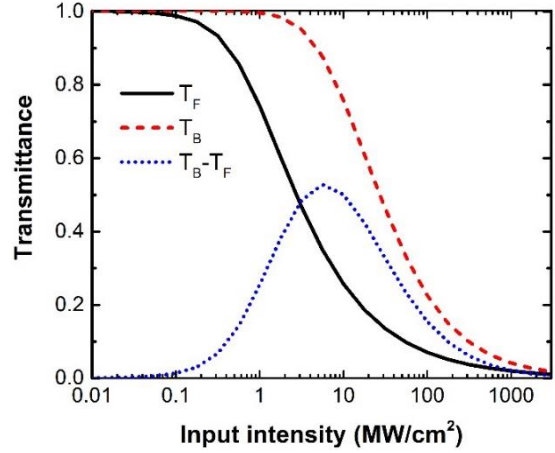

**Figure 3.** The Kerr nonlinearity effect on the forward and backward transmittance at the EP where $\lambda = \lambda_{ENZ}$. Moderate self-induced nonreciprocity is obtained for a broad input intensity range with relative low values.

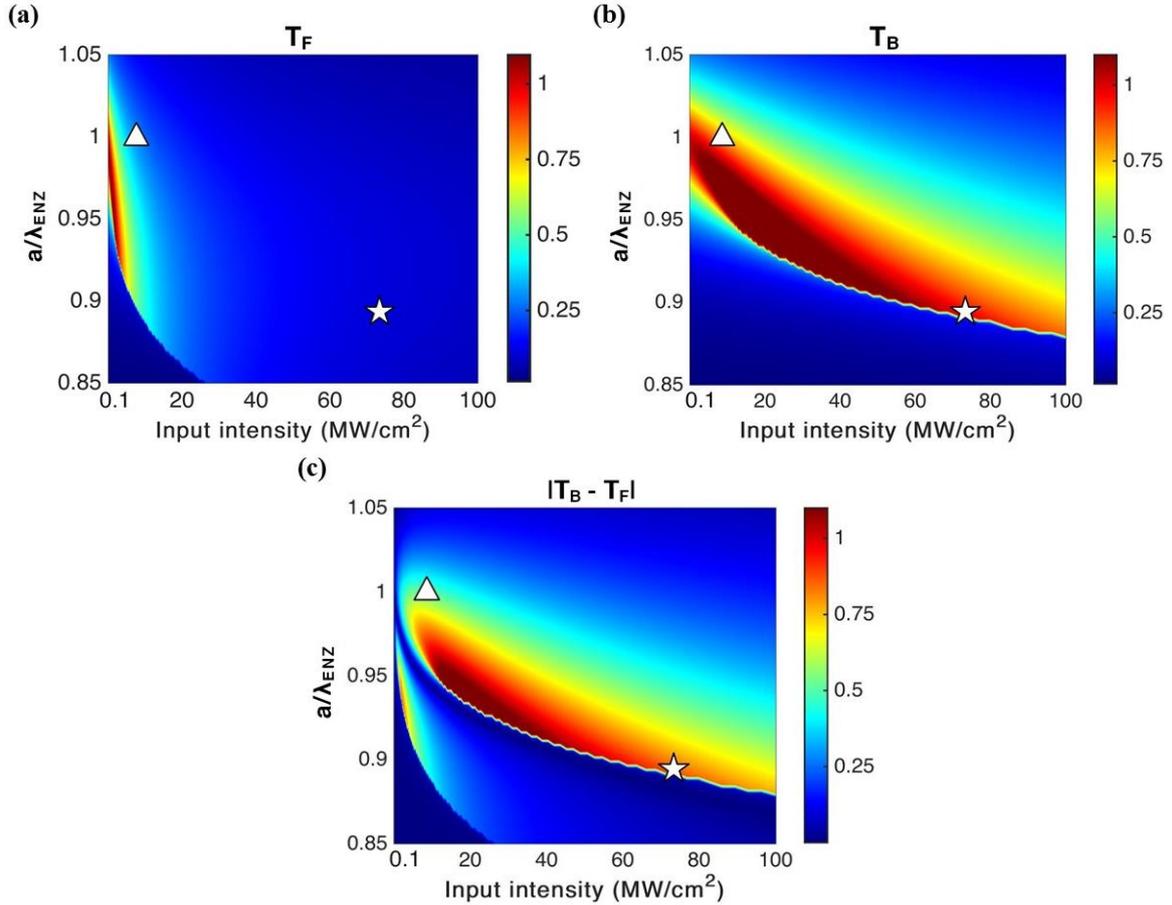

**Figure 4.** (a) Forward transmittance $T_F$, (b) backward transmittance $T_B$, and (c) transmission contrast $|T_B - T_F|$ as a function of the period $a$ and the input intensity $I_0$. The star symbol depicts an optimized point with $a = 0.891\lambda_{ENZ}$, where larger nonreciprocal transmission can be obtained compared to the EP operation marked by the triangle symbol.

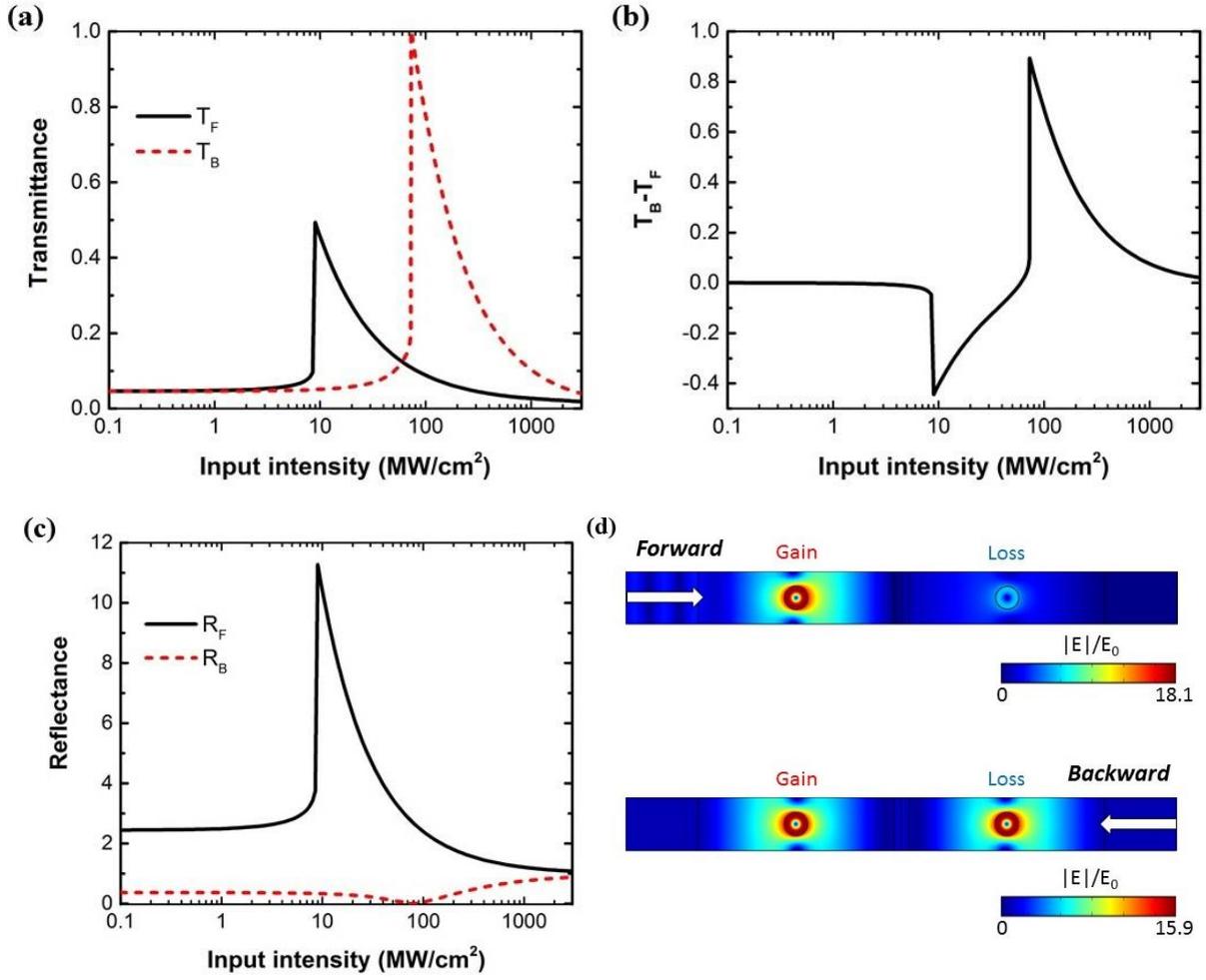

**Figure 5.** (a) Forward $T_F$ and backward $T_B$ transmittances, (b) transmission contrast $T_B - T_F$, and (c) forward $R_F$ and backward $R_B$ reflectances as functions of the input intensity when the optimized design is used $(a = 0.891\lambda_{ENZ})$. (d) The electric field enhancement distribution at $\lambda_{ENZ}$ and $I_0 = 73$ MW/cm² under forward and backward incident illumination, where $E$ is the induced electric field and $E_0$ is the electric field amplitude of the incident wave.

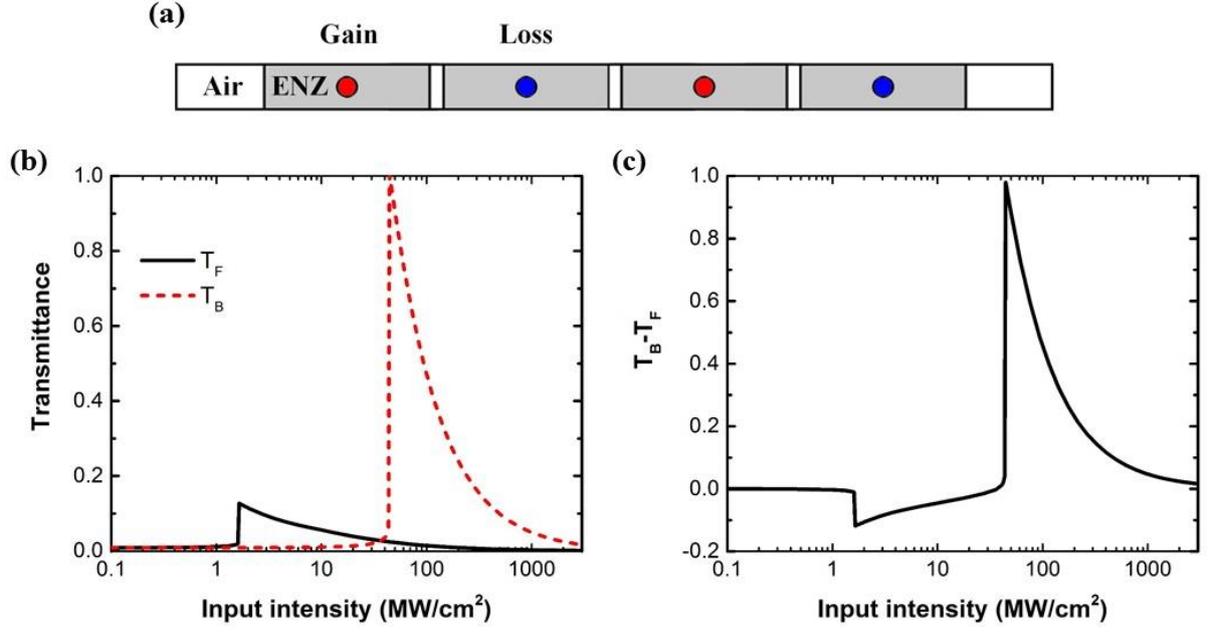

**Figure 6.** (a) Schematic of the 2-stage cascade self-induced nonreciprocal metamaterial configuration. (b) Forward $T_F$ and backward $T_B$ transmittances, and (c) transmission contrast $T_B - T_F$ as functions of the input intensity illuminating the proposed 2-stage cascade system.

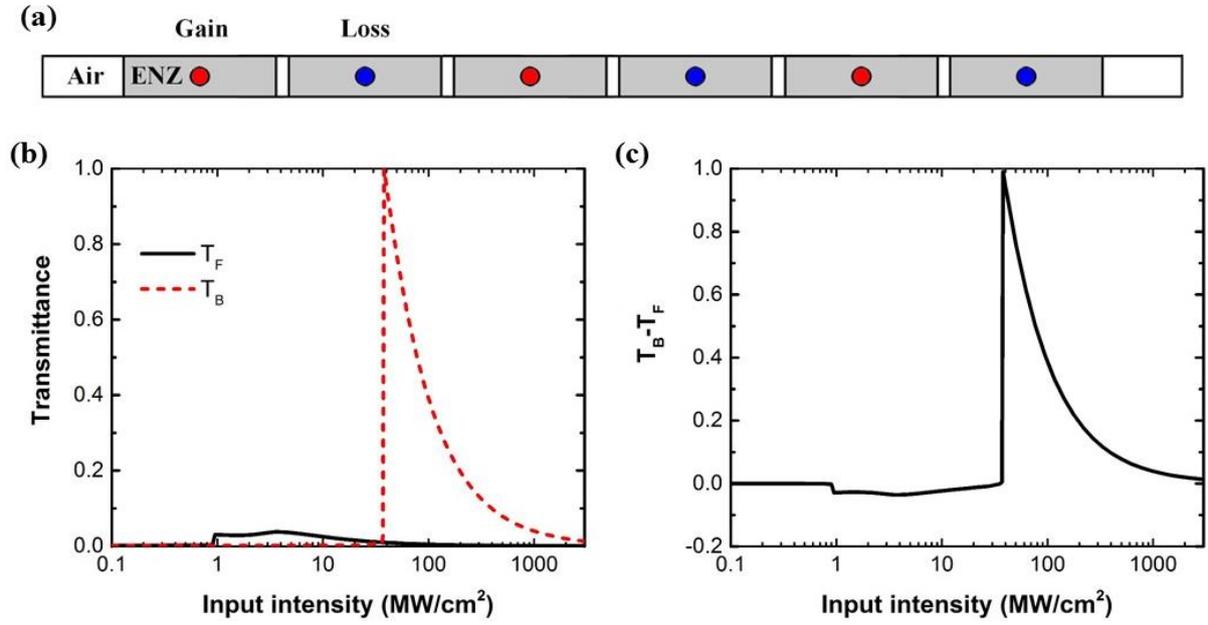

**Figure 7.** (a) Schematic of the 3-stage cascade self-induced nonreciprocal metamaterial configuration. (b) Forward $T_F$ and backward $T_B$ transmittances, and (c) transmission contrast $T_B - T_F$ as functions of the input intensity illuminating the proposed 3-stage cascade system.

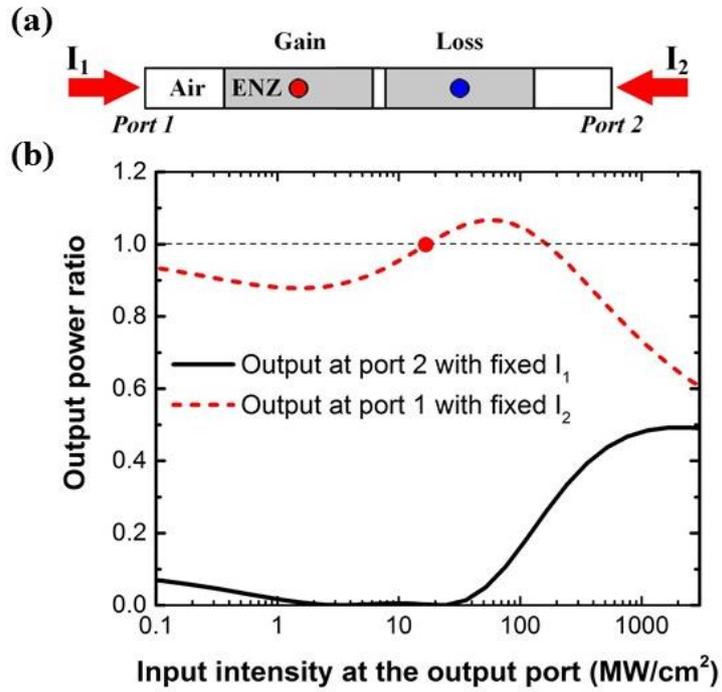

**Figure 8.** (a) Schematic of the PT-symmetric structure with two input signals simultaneously launched from the opposite ports 1 and 2. (b) The output power ratio at one port as a function of the input intensity at the same port, when the second input intensity from the opposite port is fixed to 73 MW/cm$^2$.

# Supplemental Material

# Nonreciprocal transmission in nonlinear PT-symmetric metamaterials using epsilon-near-zero media doped with defects

Boyuan Jin and Christos Argyropoulos[*]
Department of Electrical and Computer Engineering, University of Nebraska-Lincoln,
Lincoln, Nebraska 68588, USA
*christos.argyropoulos@unl.edu

1. **ENZ air gap distance *d* effect on the transmittance, reflectance, and exceptional point in the case of the linear PT-symmetric system**

The exceptional point (EP) is of great importance in the nonreciprocal functionality of the proposed nonlinear PT-symmetric metamaterial. Here, we further explore the variation of the EP as a function of the distance $d$ of the air spacer layer between the two epsilon-near-zero (ENZ) sections doped with gain/loss defects. Figure S1 presents the impact of $d$ on the forward and backward reflectance and transmittance for linear operation of the proposed PT-symmetric system by employing the same geometrical and material parameters used to obtain the results shown in Figure 2 in the main paper. The reflectance and transmittance periodically fluctuate for $d = m\lambda_{ENZ}/2$, where $m = 1, 2...$. At these points, the system exhibits a Fabry–Pérot resonant response, where the transmission/reflection and field distributions are different compared to the EP operation.[1] In addition, the gain and loss defects are embedded in only one ENZ segment when $d = 0$. The absence of the air gap results in the lack of coupling between the gain and loss materials and the EP cannot be excited for the $d = 0$ case. Interestingly, $R_B = 0$ (direct indication of an EP) only at $d = (2n-1)\lambda_{ENZ}/4$ points, where $n=1,2...$ and these distances lead to a tunable EP. We have verified that the eigenvalues coalesce only at the $R_B = 0$ points (not shown here/similar to Figure 2(b) in the main paper) and not at the Fabry–Pérot resonances. In this work, we choose to operate at the minimum distance $d = \lambda_{ENZ}/4$ (intersecting dashed lines in Figure S1(b)) to minimize the nonreciprocal device size. Interestingly, our simulations demonstrate that the EP can occur for particular distances $d$ of the air spacer layer, which is different from previous relevant works of other PT-symmetric systems,[1, 2] where the EP was found to be independent of the distance between the loss and gain parts. The impedances of the zero refractive index loss/active media in the current configuration have a reactive or imaginary part due to the stored reactive energy along their elongated thicknesses, different from the purely real impedances used in previous works.[1, 2] Hence, the formed EP can be tuned as we vary the length of the air spacer layer between the two ENZ media

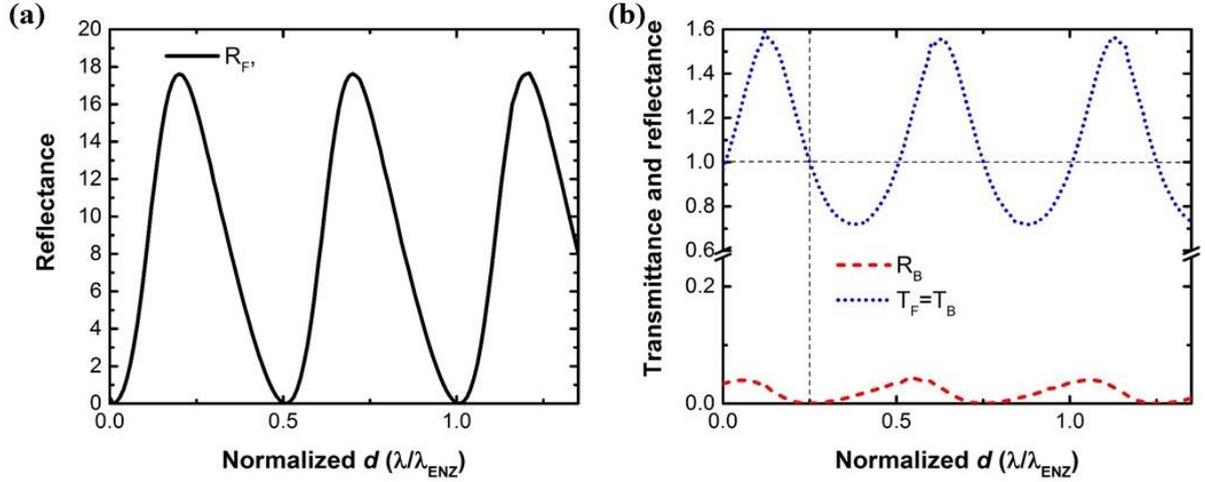

**Figure S1.** Computed (a) $R_F$ and (b) $R_B$, $T_F$ and $T_B$ of the linear PT-symmetric system as functions of $d$, i.e., the air gap distance between the two ENZ sections.

## 2. Transmittance comparison among single and cascade nonlinear PT-symmetric systems

In this section, the forward and backward transmittances of the single, 2- and 3- stage cascade systems, presented in Figures 5(a), 6(a), and 7(a) in the main paper, respectively, are compared. The computed results versus frequency are presented in the same plot shown in Figure S2. These spectra are obtained at the peak of the backward transmittances presented in Figures 5(a), 6(a), and 7(a) in the main paper. Therefore, the geometrical and material parameters are the same among the different systems except of the input intensity and periodicity $a$. The corresponding parameters are $I_0 = 73$ MW/cm$^2$ and $a = 0.891\lambda_{ENZ}$ for the single-stage configuration, and $I_0 = 44.5$ MW/cm$^2$ and $a = 0.928\lambda_{ENZ}$ or $I_0 = 38$ MW/cm$^2$ and $a = 0.938\lambda_{ENZ}$ for the 2- or 3-stage cascade systems, respectively. With the increase of the cascade stages, the forward transmittance is suppressed, as it can be clearly seen in Figure S2(a), which means that broader nonreciprocal transmission contrast is obtained by using the cascade structures. The maximum of the backward transmittance is always equal to one, independent of the used cascade or single configuration, as it is presented in Figure S2(b), a favorable condition for all-optical diode and isolator applications. The backward transmittance edge at longer wavelengths is steep, while the other edge at shorter wavelengths is more gradual. Besides, the peak of backward transmittance becomes steeper and more narrowband when more cascade stages are included, a common feature of cascading systems.

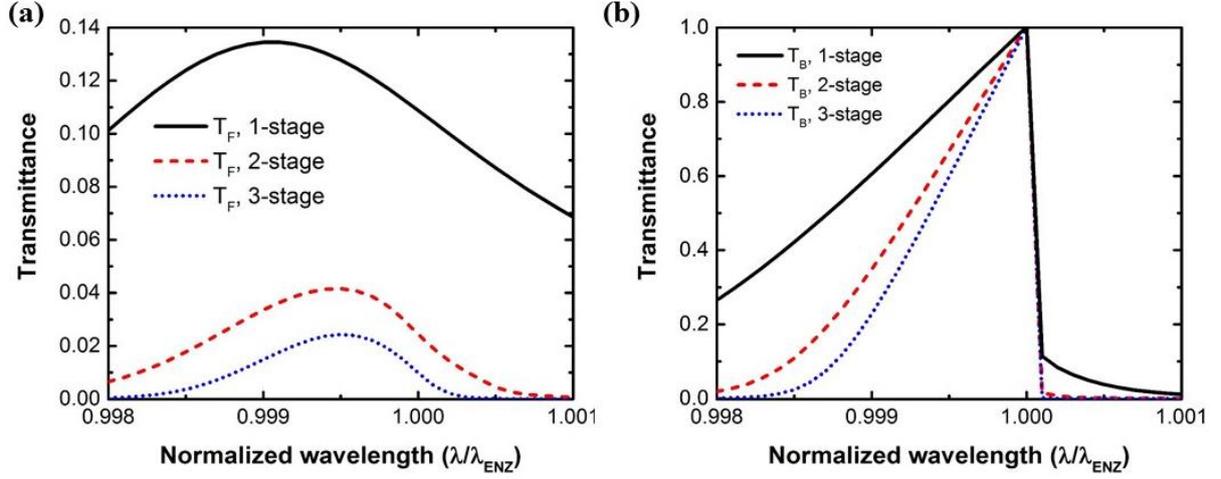

**Figure S2.** Comparison of (a) forward $T_F$ and (b) backward $T_B$ transmittances as functions of the normalized wavelength for the single, 2- and 3-stage cascade nonreciprocal systems.

The dependence of transmittance on the input intensity is compared in Figure S3 among the single, 2- and 3-stage cascade systems. The data is extracted from Figures 5(a), 6(b), and 7(b) in the main paper and the operation wavelength for these systems is fixed to $\lambda_{ENZ}$. It can be seen in Figure S3 that the required input intensity to achieve the peak transmittance is reduced for both incident directions in the cascade configurations compared to the single PT-symmetric system. By cascading more PT-symmetric units, we can observe a rapid decrease of the forward transmittance demonstrated in Figure S3(a) excited by always lower input intensity values, which combined with the unitary backward transmittance shown in Figure S3(b) will lead to a substantially improvement in the transmission contrast by using the cascade nonlinear PT-symmetric system. Further contrast improvement can be achieved by cascading more PT-symmetric metamaterial unit cells.

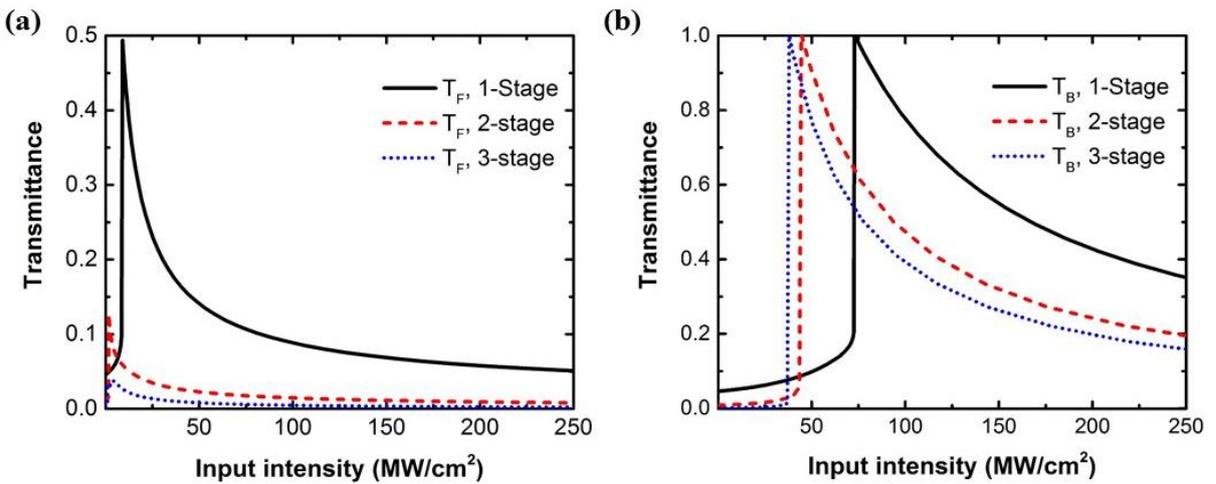

**Figure S3.** Comparison of (a) forward $T_F$ and (b) backward $T_B$ transmittances as functions of the input intensity for the single, 2- and 3-stage cascade nonreciprocal systems. The operation wavelength for these systems is fixed to $\lambda_{ENZ}$.

## 3. Electric field distribution of the 2- and 3-stage cascade nonlinear PT-symmetric systems

The spatial distribution of the electric field enhancement is critical in order to boost the strength of the Kerr nonlinear effect and the resulted nonreciprocal response. The electric field enhancement distribution for the 2- and 3-stage cascade systems is computed and shown in Figure S4. The nonlinearity of the system is considered in these plots, and the input intensity is fixed to 44.5 or 38 MW/cm$^2$ for the 2- and 3-stage cascade systems, respectively. Compared to Figure 5(d) in the main paper, it can be seen that the maximum field enhancement in both directions escalates with the increase of the cascading stages, which is another explanation of why the nonlinear nonreciprocal response can be obtained with lower input intensity values in the cascade configurations. The electric field is gradually attenuated after passing every ENZ segment in the forward propagation direction. On the contrary, the electric field enhancement distribution remains the same along each cascade stage in the backward propagation direction, leading to perfect nonreciprocal transmission.

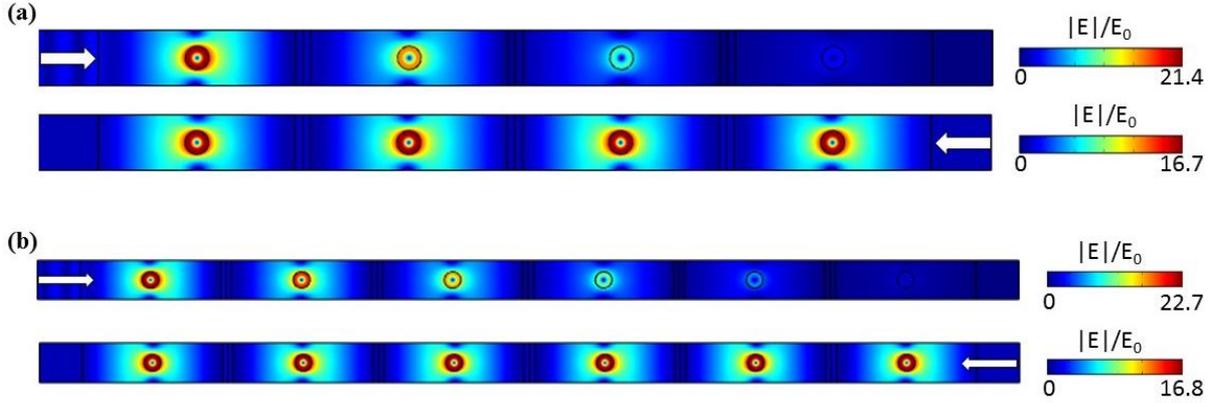

**Figure S4.** The electric field enhancement distribution at $\lambda_{ENZ}$ for (a) 2-stage and (b) 3-stage cascade systems and for forward and backward propagation direction (up and down plots of each caption, respectively), where $E$ is the induced electric field and $E_0$ is the amplitude of the input electric field. The nonlinearity is included in these simulations.

## 4. The addition of nonlinearity in the ENZ media

The nonlinear polarization is given by $P_{NL} = \varepsilon_0 \chi^{(3)} |E|^2 E$. It denotes that the Kerr effect is superlinearly dependent on the electromagnetic field. Therefore, in our manuscript, we have neglected the nonlinearity of the ENZ media, because the electric field is mainly confined in the defects of the proposed PT-symmetric system, as shown in Figure 5(d) in the main paper. However, if the ENZ media have stronger nonlinear susceptibilities $\chi^{(3)}_{ENZ}$ than that of the defects, they would also generate perceivable Kerr effect leading to a decrease in the required input intensity values to trigger the high nonreciprocal transmission contrast. This advantageous feature can decrease the system's power consumption and mitigate the power restrictions on the input pulse signal.

Here, we assume $\chi^{(3)}_{ENZ} = 1 \times 10^{-19}$ m$^2$/V$^2$, a typical nonlinear coefficient of transparent conductive oxides (TCOs), such as indium tin oxide (ITO) and aluminum zinc oxide (AZO), that can exhibit high nonlinearities at near-infrared frequencies.[3-5] Figure S5(a) shows the dependence of forward and backward transmittance on the input intensity when the nonlinearities of both ENZ media and their defects are included. The dimensions of the PT-symmetric system are the same compared to those used in Figure 5 in our manuscript, except we slightly detune the periodicity $a = 0.864\lambda_{ENZ}$ to obtain the EP and, as a result, unity transmittance. Compared to Figure 5(a), where it was assumed that $\chi^{(3)}_{ENZ} = 0$, the input intensity that can trigger strong self-induced nonreciprocal response diminishes from $I_0 = 73$ MW/cm$^2$ to $I_0 = 39$ MW/cm$^2$. At this even lower input intensity value, the system is fully transparent in the backward direction ($T_B = 1$), while the transmittance in the forward illumination is very low $T_F = 0.10$. As shown in Figure S5(b), the maximum difference in transmittance in this case is $(T_B - T_F)_{max} = 0.9$, similar to the results presented in the main paper but obtained for much lower input intensity values. Hence, interestingly, the strong self-induced nonreciprocal transmission can be preserved and even improved when the ENZ media have strong nonlinearity.

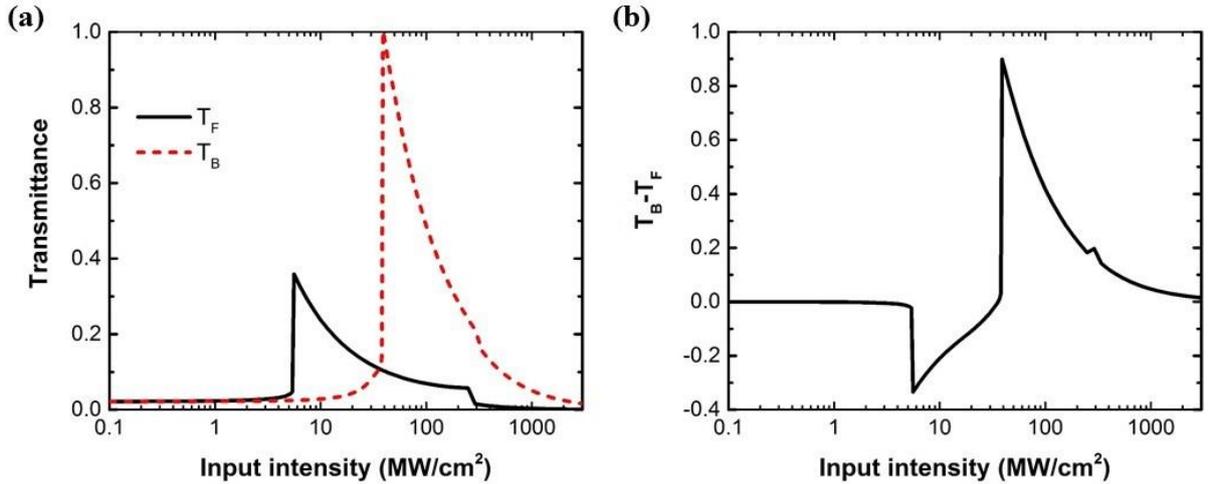

**Figure S5.** (a) Forward $T_F$ and backward $T_B$ transmittances, and (b) transmission contrast $T_B - T_F$ as functions of the input intensity when $\chi^{(3)}_{ENZ} = 1 \times 10^{-19}$ m$^2$/V$^2$. All the dimensions are the same to those in Figure 5 in the main paper except for $a = 0.864\lambda_{ENZ}$.

## 5. Lossy ENZ media

Almost lossless ENZ response, or equivalently low loss ZIM metamaterials that are needed for the currently proposed nonlinear design, can be realized by artificially engineered structures, such as waveguides operating at microwave frequencies,[6, 7] and photonic crystals (PCs) at optical frequencies.[8, 9] However, other ENZ materials, such as TCOs[3-5] and silicon carbide (SiC),[10-12] operating at near- and mid-infrared frequencies, respectively, have a moderate amount of loss due to their nonzero imaginary permittivity values.

It is noteworthy to mention that the PT-symmetric relationship $n(\vec{r}) = n^*(-\vec{r})$ will no longer be satisfied by including losses in the ENZ media to the nonreciprocal design presented in the main paper. The gain and loss will become imbalanced, since there is going to be more loss than gain in the proposed system. However, by adjusting the gain coefficient of the defect and minimizing the dimensions of the proposed structure, we can re-establish the PT-symmetry or EP formation, and, as a result, the nonreciprocal response.

Figure S6 shows the computed linear reflectance and transmittance as functions of $\delta$ (the loss/gain coefficient of the defect) under both illuminating directions. The ENZ media is assumed to be SiC in this case, where $\text{Im}(\varepsilon_{ENZ}) = -0.03$ at its mid-infrared ENZ response.[10-12] To reduce the power loss in the lossy ENZ material, the thickness of the ENZ media is decreased to $b = \lambda_{ENZ}$, leading to the requirement of larger periodicity $a$ and gain/loss value $\delta$ of the defect. The range of gain $\delta$ values used in Figure S6 are practically feasible.[13, 14] By using a larger periodicity $a = 2\lambda_{ENZ}$ and an air gap thickness $d = 0.52\lambda_{ENZ}$, it can be seen that the reflectance and transmittance can reach $T_F = T_B = 1$, $R_F = 6$, and $R_B = 0.037$ at $\delta = 0.13$ (marked with the dashed vertical lines in Figure S6). This behavior (unidirectional reflectionless transparency) can only appear at an EP, which can be used to obtain strong self-induced nonreciprocal transmission even in the case of lossy ENZ media. Hence, it is demonstrated that the proposed structure can have similar nonreciprocal dynamics when lossy ENZ media are used.

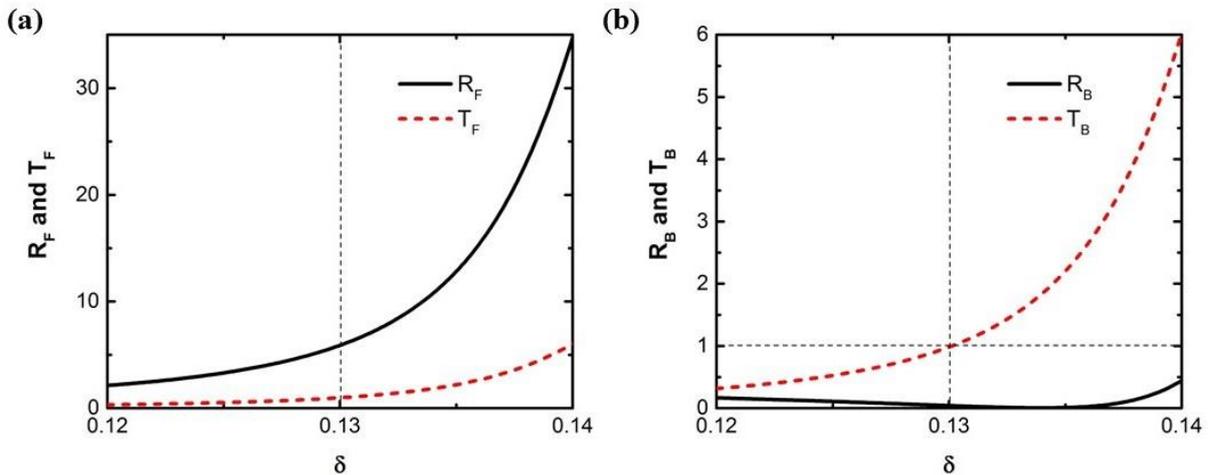

**Figure S6.** Linear transmittance $(T)$ and reflectance $(R)$ at $\lambda = \lambda_{ENZ}$ for (a) forward or (b) backward illumination as functions of $\delta$, where $b = \lambda_{ENZ}$, $a = 2\lambda_{ENZ}$, and the ENZ media is lossy with an imaginary permittivity part equal to SiC ($\text{Im}(\varepsilon_{ENZ}) = -0.03$).